\documentclass[submission,copyright,creativecommons]{eptcs}
\providecommand{\event}{ICLP Doctoral Consortium 2021} 
\usepackage{underscore}           
\usepackage{cite}
\usepackage[all]{foreign}
\usepackage{amsthm}
\usepackage[inline]{enumitem}
\usepackage{hhline}
\usepackage{float}
\usepackage{array}
\usepackage{subcaption}

\newcommand{\knobel}{knowledge/belief\xspace}

\newcommand*{\defemph}[1]{\ensuremath{\mathsf{#1}}}
\newcommand*{\agent}[1]{\ensuremath{\lowercase{\defemph{#1}}}}
\newcommand*{\opC}{\ensuremath{\mathbf{C}}}
\newcommand*{\cAlpha}[1]{\ensuremath{\opC_\alpha{#1}}}
\newcommand*{\eAlpha}[1]{\ensuremath{\mathbf{E}_\alpha{#1}}}
\newcommand*{\eAlphaIter}[2]{\ensuremath{\mathbf{E}^{#1}_\alpha{#2}}}
\newcommand*{\bB}[2]{\ensuremath{\mathbf{B}_{\defemph{#1}}{#2}}}
\newcommand*{\del}{DEL}
\newcommand*{\brel}[1]{\ensuremath{\calB_{\defemph{#1}}}}
\newcommand*{\ourL}{\mAR}
\newcommand*{\mAL}{\mbox{\ensuremath{m\mathcal{A}^*}}\xspace}

\newcommand*{\mAR}{\mbox{\ensuremath{m\mathcal{A}^\rho}}\xspace}
\newcommand*{\ck}{common knowledge}
\newcommand*{\lAG}{\ensuremath{\mathcal{L}_{\sAG}}}
\newcommand*{\lag}{\lAG}
\newcommand*{\lagC}{\ensuremath{\lAG^{\opC}}}
\newcommand*{\axT}{\textbf{T}}
\newcommand*{\axFour}{$\mathbf{4}$}
\newcommand*{\axFive}{$\mathbf{5}$}
\newcommand*{\axD}{\textbf{D}}
\newcommand*{\axK}{\textbf{K}}
\newcommand*{\logic}[3]{\textbf{#1}$_{#2}^{\mathbf{#3}}$}
\newcommand*{\state}[2]{\ensuremath{(M_{\defemph{#1}},\defemph{#2})}}
\newcommand*{\sAG}{\ensuremath{\mathcal{AG}}}

\newcommand*{\sF}{\ensuremath{\mathcal{F}}}

\newcommand*{\func}[3]{#1: #2 \mapsto #3}
\newcommand*{\implies}{\ensuremath{\Rightarrow}}
\newcommand*{\bra}[1]{\ensuremath{\{#1\}}}
\newcommand*{\tuple}[1]{\ensuremath{\langle #1 \rangle}}
\newcommand*{\calB}{\ensuremath{\mathcal{B}}}

\theoremstyle{definition}
\newtheorem{definition}{Definition}[section]
\newtheorem{example}{Example}[section]
\title{Comprehensive Multi-Agent Epistemic Planning}
\author{Francesco Fabiano
\institute{Dipartimento DMIF, Universit\`a di Udine, Udine, Italy}
\email{francesco.fabiano@uniud.it}}

\def\titlerunning{Comprehensive Multi-Agent Epistemic Planning}
\def\authorrunning{F. Fabiano}

\begin{document}

	\maketitle              
\begin{abstract}
	
	Over the last few years, the concept of Artificial Intelligence has become central in different tasks concerning both our daily life and several working scenarios.
	Among these tasks \emph{automated planning} has always been central in the AI research community.
	In particular, this manuscript is focused on a specialized kind of planning known as \emph{Multi-agent Epistemic Planning} (MEP).
	
	\emph{Epistemic Planning} (EP) refers to an automated planning setting where the agent reasons in the space of knowledge/beliefs states and tries to find a plan to reach a desirable state from a starting one. Its general form, the MEP problem, involves multiple agents who need to reason about both the state of the world and the information flows between agents.
	To tackle the MEP problem several tools have been developed and, while the diversity of approaches has led to a deeper understanding of the problem space, each proposed tool lacks some abilities and does not allow for a comprehensive investigation of the information flows.
	That is why, the objective of our work is to formalize an environment 
	where a complete characterization of the agents' knowledge/beliefs interaction and update is possible.
	In particular, we aim to achieve such goal by defining a new action-based language for \emph{multi-agent epistemic planning} 
	and to implement an epistemic planner based on it.
	This solver should provide a tool flexible enough to reason on different domains, \eg economy, security, justice and politics,
	where considering others' knowledge/beliefs could lead to winning strategies.
	
\end{abstract}
\section{Introduction}
The proliferation of agent-based and IoT technologies has enabled the development of novel applications involving hundreds of agents. 
To operate in such domains agents need to make decisions based on both the available problem information and on their knowledge, or belief, about the preferences and actions of other agents involved in, or impacted by, their decisions.

To maximize the potentials of such autonomous systems, \emph{multi-agent planning} and scheduling research \cite{dovier1,de2009introduction,de2003resource,lipovetzky2017best,richter2010lama} will need to keep pace. 
As already said, creating a plan for multiple agents to achieve a goal will need to take into consideration agents' knowledge, beliefs and to account for aspects like trust, dishonesty, deception and incomplete knowledge.
The planning problem in this new setting is referred to as \emph{Multi-agent Epistemic Planning} (MEP) in the literature. 
MEP is a generalization of the concept of \emph{Epistemic Planning} (EP),
which refers to a single agent setting that reasons about knowledge and beliefs. In the MEP scenario, agents reason about the information that other agents have of each other while also considering how such information is shared. That is, MEP is the research area that attempts to create artificial agents that build plans considering the information flows between the agents themselves.

Nevertheless, reasoning about knowledge and beliefs is not as direct as reasoning on the \textquotedblleft physical" state of the world.
In fact, expressing belief relations between a group of agents often implies to consider \emph{nested} and \emph{group} beliefs that are not easily extracted from the state description by a human reader. For this reason, it is necessary to develop a complete and accessible action language to model multi-agent epistemic domains~\cite{baral2015action} and to advance also in the study of epistemic solvers~\cite{le2018efp,bolander2011epistemic,wan2015complete,baral2015action,muise2015planning,huang2017general,icaps20}.
While these solvers may have different restrictions, \eg considering only the concept of \emph{knowledge} while not allowing the idea of \emph{belief} or not allowing for \emph{dynamic common \knobel}, all of them share the same goal: to plan while considering the information flows between the agents.

In our research we are exploring the epistemic planning problem with particular focus on
\begin{enumerate*}[label=\roman*)]
	\item formalizing an updated version of the action languages \ourL~\cite{icaps20} and E-PDDL~\cite{epddl}; and 
	\item on developing a comprehensive multi-agent epistemic solver flexible enough to be used in several scenarios such as: economy, security, justice and politics.
\end{enumerate*} 
\section{Background}\label{sec:background}
This section mostly stems from~\cite{DBLP:journals/corr/abs-1909-08259} where the same concepts where introduced in order to describe the foundations of the work presented in this manuscript.

\subsection{Dynamic Epistemic Logic}
Logicians have always been interested in describing \emph{the state of the world} through formalism that would allow  to reason on the world with logic itself.
This interest has lead, among other things, to the formalization of the well-known planning problem~\cite{modernApproach} and to the introduction of several \emph{modal logics}~\cite{van2007dynamic,Chagrov1997,smullyan2012first} used to describe different types of domains.
\emph{Dynamic Epistemic Logic} (\del), in particular, is used to reason not only on the state of the world but also on \emph{information change}.
As said by Van Ditmarsch \textit{et al.}~\cite{van2007dynamic}: ``\emph{information} is something that is relative to a subject who has a certain perspective on the world, called an \emph{agent}, and that is meaningful as a whole, not just loose bits and pieces. This makes us call it \emph{knowledge} and, to a lesser extent, \emph{belief}".
The idea behind \del\ is, therefore, to have a formalization that allows to reason on dynamic domains where, not only the state world is taken into consideration, but also the knowledge/beliefs that the agents have about the world and about the knowledge/beliefs of each other are considered.

\subsubsection{Epistemic Logic}
Dynamic Epistemic Logic is, clearly, connected to \emph{epistemic logic}: that is the logic that allows to reason on the knowledge/belief of agents in static domains.
This logic is based on two main concepts:
\begin{enumerate*}[label=\roman*)]
	\item \emph{Kripke structures}, a data structure that is widely use in literature~\cite{fagin1994reasoning,baral2015action,van2007dynamic} to model its semantics; and
	\item \emph{belief formulae}, a type of formula that takes into consideration epistemic operators and is used to represent the knowledge/beliefs of the agents.
\end{enumerate*}
As it is beyond the scope of this work to give an exhaustive introduction on epistemic logic let us provide only the fundamental definitions and intuitions; the reader who has interest in a more detailed description can refer to~\cite{fagin1994reasoning}.

Let \sAG\ be a set of  agents and let \sF\ be a set of propositional variables, called \emph{fluents}.
We have that each \emph{world} is described by a
subset of elements of \sF\ (intuitively, those that are \textquotedblleft true" in the world).
Moreover, in epistemic logic each agent \agent{ag} $\in \sAG$ is associated with an epistemic modal operator \bB{ag}{} 
that intuitively represents the knowledge/belief of \agent{ag}.
Finally, epistemic \emph{group operators} \eAlpha\ and \cAlpha\ are also introduced in epistemic logic. Intuitively \eAlpha\ and \cAlpha\ represent the knowledge/belief of a group of agents $\alpha$ and
the \emph{\ck/belief} of $\alpha$ respectively.
To be more precise, following the work of Baral \textit{et al.}~\cite{baral2015action}, we have that:
\begin{definition}
	A \emph{fluent formula} is a propositional formula built using the propositional
	variables in \sF\ and the traditional propositional operators
	$\wedge,\vee,\implies,\neg$. We make use $\top$ and $\bot$ to indicate
	\emph{True} and \emph{False}, respectively.
	A \emph{fluent atom} is a formula composed by just an element $\defemph{f} \in \sF$, instead
	a \emph{fluent literal} is either a fluent atom $\defemph{f} \in \sF$ or its negation $\neg \defemph{f}$. During this work let us refer to fluent literals simply as \emph{fluents}.
\end{definition}

\begin{definition}
	A \emph{belief formula} is defined as follow:
	\begin{itemize}
		\item A fluent formula is a belief formula;
		\item let $\varphi$ be belief formula and $\agent{ag} \in \sAG$, then  $\bB{ag}{\varphi}$ is a belief
		formula;
		\item let $\varphi_1, \varphi_2$ and $\varphi_3$ be belief formulae,
		then $\neg \varphi_3$ and $\varphi_1 \,\mathtt{op}\, \varphi_2$ are belief
		formulae, where $\mathtt{op} \in \bra{\wedge,\vee, \implies}$;
		\item all the formulae of the form \eAlpha{\varphi} or \cAlpha{\varphi}
		are belief formulae, where $\varphi$ is itself a belief formula and
		$\emptyset \neq \alpha \subseteq \sAG$.
	\end{itemize}
\end{definition}
From now on let us denote with \lagC\ the language of the belief formulae over the
sets $\sF$ and $\sAG$ and with \lag\ as the language over beliefs formulae that
does not allow the use of   \opC.

\begin{example}
	Let us consider the formula $\bB{ag_1}{\bB{ag_2}{\varphi}}$. This formula expresses that
	the agent \agent{ag_1} believes that the agent \agent{ag_2} believes that $\varphi$ is true, instead,
	$\bB{ag_1}\neg \varphi$ expresses that the agent \agent{ag_1} believes that $\varphi$ is false.
\end{example}
As mentioned above the classical way of providing a semantics for the language of epistemic logic is in terms
of \emph{pointed Kripke structure}~\cite{Kripke1963-KRISCO}. More formally:

\begin{definition}
	A \emph{Kripke structure} is a tuple \tuple{S, \pi, \brel{1},$\dots$ , \brel{n}}, such that:
	\begin{itemize}
		\item S is a set of worlds;
		\item $\func{\pi}{S}{2^{\sF}}$ is a function that associates an interpretation
		of \sF\ to each element of S; 
		\item for $1 \leq \defemph{i} \leq \defemph{n}$, $\brel{i} \subseteq S \times S$  is a binary relation over S.
	\end{itemize}
\end{definition}
\begin{definition}
	A \emph{pointed Kripke structure} is a pair \state{}{s} where $M$ is a Kripke structure
	as defined above and $\defemph{s} \in S$, where $\defemph{s}$ represents the real world.
\end{definition}
Following the notation of~\cite{baral2015action}, allow us to indicate with
$M[S], M[\pi]$ and $M[\defemph{i}]$ the components $S,\pi$ and $\brel{i}$ of $M$,
respectively.
Intuitively  $M[S]$ captures all the worlds that the agents believe to be possible and $M[\defemph{i}]$ encodes the beliefs of each agent. More formally the semantics on pointed Kripke structure is as follows:

\begin{definition}
	Given the belief formulae
	$\varphi,\varphi_{1},\varphi_{2}$, an agent \agent{ag_i}, a group of agents $\alpha$, a pointed Kripke structure ($M = \tuple{S, \pi, \brel{1}, ..., \brel{n}}$, \defemph{s}):
	\begin{enumerate}[label= \emph{(}\roman*\emph{)}]
		\item $\state{}{s} \models \varphi$ if $\varphi$ is a fluent formula and $\pi(\defemph{s})
		\models \varphi$;
		\item $\state{}{s} \models \bB{ag_i}{\varphi}$ if for each \defemph{t} such that
		$\defemph{(s,t)} \in \brel{i}$ it holds that $ \state{}{t} \models \varphi$;
		\item $\state{}{s} \models \eAlpha{\varphi}$ if $\state{}{s} \models \bB{ag_i}{\varphi}$ for all \agent{ag_i} $\in \alpha$;
		\item $\state{}{s} \models \cAlpha{\varphi}$ if
		$\state{}{s} \models \eAlphaIter{k}{\varphi}$ for every
		$k\geq0$, where $\eAlphaIter{0}{\varphi} = \varphi$ and
		$\eAlphaIter{k+1}{\varphi} =\eAlpha{(\eAlphaIter{k}{\varphi})}$;
		\item the semantics of the traditional propositional operators is as usual.
	\end{enumerate}
\end{definition}

\subsubsection{Knowledge or Belief}

As pointed out in the previous paragraph the modal operator $\bB{ag}{}$ represents the worlds' relation in a Kripke structure 
and, as expected,	different relations' properties imply different meaning for $\bB{ag}{}$.
In particular in this work we are interested in representing the knowledge or the beliefs of the agents.
The problem of formalizing these two concepts has been studied in depth bringing to an accepted formalization for both knowledge and beliefs.
In particular, when a relation\footnote{In our case the relation between the world in a Kripke structure.} respects all the axioms presented in Table~\ref{tab:axioms} is called an \emph{\textbf{S5}} relation and it encodes the concept of knowledge while when it encodes all the axioms but \axT\ it characterizes the concept of belief.
Following this characterization we refer to knowledge and belief as \textbf{S5} and \textbf{KD45} logic respectively.

\begin{table}[h]
	\centering
	\begin{tabular}{||c||c||}
		\hhline{|t:=:t:=:t|}
		\multicolumn{1}{||c||}{\phantom{...}\textbf{Property of $\brel{}$}\phantom{...}}
		& \multicolumn{1}{c||}{\phantom{...}\textbf{Axiom}\phantom{...}}\\
		\hhline{|:=::=:|}
		\multicolumn{1}{||l||}{$\calB_{\mathtt{i}}\varphi  \implies \varphi$}
		& \multicolumn{1}{c||}{\axT}\\
		\hhline{||-||-||}
		\multicolumn{1}{||l||}{$\calB_{\mathtt{i}}\varphi  \implies \calB_{\mathtt{i}}\calB_{\mathtt{i}}\varphi$}
		& \multicolumn{1}{c||}{\axFour}\\
		\hhline{||-||-||}
		\multicolumn{1}{||l||}{$\neg \calB_{\mathtt{i}}\varphi  \implies \calB_{\mathtt{i}}\neg \calB_{\mathtt{i}}\varphi$}
		& \multicolumn{1}{c||}{\axFive}\\
		\hhline{||-||-||}
		\multicolumn{1}{||l||}{$\neg \calB_{\mathtt{i}} \bot$}
		& \multicolumn{1}{c||}{\axD}\\
		\hhline{||-||-||}
		\multicolumn{1}{||l||}{$(\calB_{\mathtt{i}}\varphi \wedge \calB_{\mathtt{i}}(\varphi \implies \psi)) \implies  \calB_{\mathtt{i}}\psi$}
		& \multicolumn{1}{c||}{\axK}\\
		\hhline{|b:=:b:=:b|}
	\end{tabular}
	\caption{\label{tab:axioms}Knowledge and beliefs axioms.~\cite{fagin1994reasoning}.}
\end{table}
Intuitively the difference between the two logics is that an agent cannot \textit{know} something that is not true in \textbf{S5} but she can \textit{believe} it in \textbf{KD45}.
As this introduction is not supposed to explore in depth this topic we will not go into further detail, and we address the interested reader to~\cite{fagin1994reasoning}.

\subsection{Multi-Agent Epistemic Planning}
\emph{Epistemic planning}~\cite{bolander2011epistemic} refers to the generation of plans for multiple agents to achieve goals which can refer to the state of the world, the beliefs of agents and/or the knowledge of agents. It has recently attracted the attention of researchers from various communities such as planning, dynamic epistemic logic and knowledge representation.

With the introduction of the classical planning problem, in the early days of artificial intelligence, several action languages (e.g., $\mathcal{A}$, $\mathcal{B}$ and $\mathcal{C}$) have been developed~\cite{gelfond1998action} and have also provided the foundations for several successful approaches to automated planning.
However, the main focus of these research efforts has been about reasoning within single agent domains. 
In single agent domains reasoning about information change mainly involves reasoning about what the agent knows about the world and how she can manipulate it to reach particular states.
In multi-agent domains, on the other hand, an agent's action may change the world, other agents' knowledge about it and their knowledge about other agents' knowledge about the world. Similarly, goals of an agent in a multi-agent domain may involve manipulating the knowledge of other agents---in particular, this may concern not just their knowledge about the world, but also their knowledge about other agents' knowledge about the world.

As said before, epistemic planning is not only interested in the state of the world but also in the beliefs and the knowledge of the agents. Although there is a large body of research on multi-agent planning~\cite{de2003resource,de2009introduction,goldman2004decentralized}, very few efforts address the above mentioned aspects of multi-agent domains, which pose a number of new research challenges in representing and reasoning about actions and change.

Due to its complexity, the majority of search based epistemic planners (e.g.,~\cite{crosby2014single,engesser2017cooperative,kominis2017multiagent, huang2017general,muise2015planning, wan2015complete}) impose certain restrictions, such as the finiteness of the levels of nested beliefs. 
Such restrictions permit, for instance, in~\cite{muise2015planning, huang2017general, wan2015complete} to solve the problem by translating it into classical planning.
Moreover,  existing epistemic action languages~\cite{bolander2011epistemic,muise2015planning,baral2015action,icaps20,iclp20} are able to model several families of problems,
but cannot comprehensively reason on aspects like trust, dishonesty, deception and incomplete knowledge.
In order to exploit epistemic reasoning in complex real-world scenarios it is paramount to define formalism that are able to consider the aforementioned concepts.

The planners~\cite{le2018efp,icaps20,iclp20} and the language \mAL, that is the language that implemented by the these solvers, are the starting points of this research work.
As we will explain in the following sections the main objectives of this thesis are:
\begin{enumerate*}[label=\roman*)]
	\item to formalize a complete and flexible epistemic action-based language;
	\item to study alternative representations for epistemic models; and
	\item to implement a competitive planner to reason on information change.
\end{enumerate*} 
Given the amount of information to properly introduce \mAL\ and the concept of \emph{event update semantics} we redirect the reader who is not familiar with these two topics to~\cite{baral2015action} for a clear and complete explanation.
\section{Goals of the research}\label{sec:goals}
The presented research aims to develop an epistemic action language and an epistemic solver general enough that could be adopted to explore the idea of comprehensive epistemic planning.
In particular, this work aims to develop an environment in which it is possible to define custom \emph{update models} and, therefore, \emph{custom action types} in order to reason on all the subtle concepts related to the epistemology: \eg trust, misconception, lies and so on.

Some existing epistemic solvers~\cite{le2018efp,icaps20} strictly relate action types to their underlying action language and, therefore, ``statically'' define the epistemic states update for each action type (\ie \emph{ontic}, \emph{sensing} and \emph{announcement}).
These languages define precisely how an epistemic state (e-state) is updated after the execution of an action by defining how the  e-states representation (\ie \emph{Kripke structure} or \emph{possibilities}) is structurally updated.
This update is done thanks to concepts of \emph{agents' observability} that allows to define the degree of change in the agents' beliefs w.r.t. the properties of the world modified by the action.
The newly generated e-state will then represents the ``new" knowledge/beliefs graph that will allow checking belief formulae entailment without limitations (\eg on the depth of the formulae).
This approach permits to reason on the full extent of \logic{S5}{n}{C} and \logic{KD45}{n}{C} (with $n \geq 1$) but imposes limits on what an agent can do by allowing only predefined action types.

Other approaches, \eg~\cite{muise2015planning}, are not based on structural e-state update and allow each action to define its effects explicitly.
This means that each action needs to completely specify how the agents' beliefs are updated.
While having explicit updates allows to freely characterize each action's effects, we feel that leaving such detail as input is partially in contrast with the idea of autonomous epistemic reasoning. 
Furthermore, it is easy to inadvertently omit or mis-characterize ``intricate'' \knobel relations that should instead be handled by the autonomous solving process.
Chains of \knobel derived by different degree of observability and, as well as other complex \knobel structures, may be easily mis-represented given their complex and not intuitive nature.

It is our objective to define an epistemic environment in which it is possible to define custom action types without having to rely on manually entering all the intricate \knobel relations. 

To be more precise about our research goals let us now present the major sub-goals of our work separately:
\begin{itemize}[leftmargin=*]
	\item To formalize a general and unified  \textbf{action-based language}  for epistemic domains~\cite{epddl}.
	The diversification of approaches in solving this problem certainly brings benefits to the entire community of multi-agent epistemic planning.
	Nonetheless, with many specialized planners suitable for specific scenarios, the lack of a standardized way to define input problems has led to the creation of ad-hoc languages (each one with non-transferable feature, \eg explicit vs. implicit belief update) that can become an obstacle when comparing, combining (i.e., ensemble) or testing different planners.
	We believe that a unified way of expressing MEP problems would provide researchers with a faster and less error-prone way of defining standardized benchmarks and would also relieve them from the burden of having to learn a new language each time they need to make use of a different tool.
	
	\item To define an \textbf{efficient e-state underlying data structure} and to formalize a \textbf{general e-state update} that is able to capture all the subtle possible variations of action types in epistemic reasoning.
	In particular, we envisioned two possible solutions:
	\begin{enumerate*}[label=\roman*)]
		\item to enhance the actions types presented in by Fabiano \textit{et al.}~\cite{icaps20} so that they could capture the ideas of an \emph{agent's attitude} and, therefore, reason on ideas such trust and misconception; and
		\item to allow the user to define custom update models (and custom observability groups) so that each action could be associated to a specific way of structurally modifying the e-state.
	\end{enumerate*} 
	In particular, the latter solution seems the one that is more general and also more in line with the epistemic community.
	
	\item Finally, the most important contribution of this thesis is the implementation of an \textbf{general and comprehensive epistemic solver} that can be adopted by the community as a basis for future researches.
	This planner will be based on the PDDL-like epistemic action language so that researchers from other planing community can better understand how to define epistemic problems.
	Moreover, with the introduction of agents' attitudes and custom event model the planner will allow the users to tailor action in whichever fashion they prefer without having to worry about tedious and intricate effects definitions.
	This will help in formalizing new tools in which agents are able to reason while considering \knobel relations with concepts such as lies, misconceptions, trust and so on. It would even possible to define special groups of agents that reacts differently to the actions.
	
	Whilst the generality of the planner is of the utmost importance, reducing the search times, given the inherent complexity of DEL, is also a feature that is essential to our solver.
	That is why, along with formalization of more efficient data structures and processes of e-state updates we also decided to focus on developing some domain-independent heuristics.
	These heuristics range from very simple ones, \eg the number of satisfied sub-goals, to more complex.
	An example of the latter is the updated version of the epistemic planning graph, that stems from a combination of~\cite{le2018efp} and~\cite{iclp20}.
	
	Furthermore, this planner will be accompanied by a simple user interface in which the user will be able to select between several configurations; \eg whether to employ e-state reduction, to decrease generality in favor of performances, to print out the graphic representation of the planning process or even to select between different e-states representations.
	The interface will be further enriched with a \emph{portfolio algorithm-like} procedure that will help the user to select the best planner configuration given a certain problem or even among the planner itself and some other solvers derived from the literature that may be more efficient in certain situations.
	In particular, the selection between the diverse planners, will be done following the ideas of~\cite{booch2020thinking} where meta-cognitive architectures are employed to select the best way of solving specific problems.

\end{itemize}
\section{Status of the research}\label{sec:current}
In this section we quickly present the work done on the goals introduced previously.
Let us now provide, without tedious technical details, the status of the research of the goals presented in Section~\ref{sec:goals};
the reader who is interested in a more complete introduction to these is mainly addressed to~\cite{le2018efp,cilc19Epistemic,icaps20,iclp20,booch2020thinking,epddl}.

\begin{itemize}[leftmargin=*]
	\item \textbf{Language formalization.} Considering that the main objective of this thesis is to provide a general and comprehensive epistemic solver, the first step of this process is to define an action language that will be the base for such solver.
	The formalization of this language followed diverse directions.
	Let us now briefly explain the various processes that allowed us to expand the language and to define a general and ``standard" epistemic language.
	\begin{itemize}[leftmargin=*]
		\item The initial step of this research line was to adapt the language \mAL~\cite{baral2015action} so that it can be adopted also on a data structure that differs from the one considered by Baral \textit{et al.}.
		In particular, while \mAL was defined on Kripke structures, we devised \ourL~\cite{icaps20} that works on Possibilities~\cite{Gerbrandy1997}.
		\item Following we devised for both \mAL and \ourL the concept of \emph{agents' attitudes}.
		Thanks to this formalism we were able to add variations of the epistemic actions (in a paper under blind review) where the idea of trust and misconception is considered.
		This resulted in more complicated action types that allow to describe more realistic scenarios.
		Nevertheless, these actions types were still static and predetermined making the language incapable of handling all the nuances that naturally arise when concepts such as trust and lies are considered.
		\item While trying to make the language more general we also encountered the problem of having a standardized way of defining domains across the epistemic community. We, therefore, decided to construct an input language for epistemic domains that could accommodate all the needs of the epistemic planning community while inheriting the syntax from the well known PDDL.
		From these considerations we created E-PDDL~\cite{epddl}, an adaptation of PDDL for epistemic porblems. We also created a parser~\cite{epddl} that is able to read a input in E-PDDL and convert such input into a format that the two state-of-the-art epistemic planners, \ie~\cite{muise2015planning,icaps20}, can ``understand''.
		\item As a final step we envisioned a way of defining \emph{custom update models} for both Kripke structure and Possibilities.
		These custom update models will allow the user to specify all sorts of behaviors for the actions making it possible to capture all the variations in the \knobel update in epistemology.
		In particular, this level of customization will allow to confront several theories on how the agents' \knobel must be updated when, for example, in presence of lies, stubbornness, trust ignorance and so on.
		The specification of custom update models will also be made through a PDDL-like syntax.
	\end{itemize}
	
	\item \textbf{Alternative e-state representation.} Kripke structure is the classical way to provide semantics for epistemic logic~\cite{fagin1994reasoning} but this does not mean that is the only suitable representation.
	In particular, in~\cite{icaps20} we introduced a refactor of \mAL, called \ourL, that bases the state representation on \emph{possibilities}, a data structure introduced in~\cite{Gerbrandy1997} and derived from \emph{non-well-founded\ sets} theory.
	These structures allowed the planner to have an increase in terms of performances especially thanks to the optimized transition function.
	While this transition function is not general enough to accommodate the general case (\ie when action types are not statically defined) it still has good performances on the case when the needed action types are the ``classical" ones; \ie ontic, sensing and announcement.
	Given the performance boost seen in~\cite{icaps20}, we decided to use possibilities as the main representation for e-states in the solver while maintaining also Kripke structures as it is the standard representation.
	
	\item \textbf{Epistemic Solver:} as main objective of our research we aim to provide a multi-agent epistemic planner flexible enough to be used in real world scenarios.
	This solver stems from the one firstly presented in~\cite{le2018efp}.
	During the Ph.D. period the planner has been constantly modified in order to increase its performances and to include the new studied ideas. 
	At this point of our study we have a modular C\texttt{++} solver that:
	\begin{itemize}[leftmargin=*]
		\item is able to ``understand" input specification expressed in both \ourL and E-PDDL;
		\item can choose between two epistemic state representations, \ie Kripke structure or Possibilities;
		\item is able to adopt three different transition functions for each state representation:
		\begin{enumerate*}[label=\roman*)]
			\item the one that only allows for the standard epistemic actions~\cite{baral2015action,icaps20};
			\item the update that considers the agents' attitudes and is able to consider ideas such as trust and lies; and
			\item the configuration where the action types can be defined by the user through custom update models.
		\end{enumerate*}
		These configurations are all available as they represent different trade-offs between generality and performances;
		\item allows the use of heuristics. In particular, a refined version of the epistemic planning graph presented in~\cite{le2018efp} is under development. This planning graph will also be able to adapt to the selected type of transition function and it is independent of the chosen e-state representation making it available for every planner configuration.
	\end{itemize}
	Finally, a version in Answer Set Programming (ASP) of the planner in~\cite{icaps20} has been implemented and presented in~\cite{iclp20}. This planner has the same abilities as the one in~\cite{icaps20} but thanks to its declarative nature it is more accessible to non-experts and easier to maintain. We are currently working in order to add the more general transition function also in the ASP solver. Both the versions ( C\texttt{++} and ASP) of the solver are available upon request.
\end{itemize}
%
%
%
%
%
%
\section{Open issues} \label{sec:issues}
Multi-agent epistemic planning is a relatively new field of research in computer science.
While the community is evolving rapidly there are still critical points that need to be addressed.
In what follows we summarize what, we think, are the most critical issues that need to be tackled in epistemic planning.  

\begin{itemize}[leftmargin=*]
	\item The lack of real-world examples where epistemic planning should be adopted.
	Multi-agent planners are very powerful tools able to reason on multi-agent scenarios, which are found everywhere in the real world.
	Nevertheless, the MEP community is missing strong examples that can be used as fundamental motivation of why such technology must be investigated.
	This lack of examples is probably due to the fact that epistemic reasoners do not scale well yet and, therefore, it is difficult to solve complex multi-agent problems when even toys examples are an obstacle.
	Another motivation for the absence of real-world inspired domains may be that each epistemic planner adopted its own specification language often making it difficult for other researches to adopt these tools in real tasks.
	
	\item As just mentioned another important issues is the poor scalability of the epistemic reasoners.
	Being the community relatively new it is normal that most of the effort is put in investingating the foundation of the problem rather that to optimize what already exists.
	Nonetheless, having tools that, most of the time, have not acceptable performances (w.r.t. classical planing, for example) limits the proliferation of the solvers themselves.
	It is paramount, in our opinion, to focus on the optimization of existing tools and formalism in order to have competitive epistemic reasoners that can be employed by other researchers or even in real-world scenarios. This would allow the community to gain more momentum and to grow even faster.
	Such optimizations could derive from several directions: implementation of heuristics, use of parallelism, adoption of symbolic e-state representations, combination of existing approach in a portfolio-like architecture and so on. 
\end{itemize}
%
%
%
%
%
%
\section{Acknowledgments}
This  research  is  partially  supported  by:  the  NSF  HRD 1914635  and  NSF  HRD  1345232  
projects,  the  University of  Udine PRID  ENCASE project and  the  GNCS-INdAM
2017–2020 projects.

\bibliographystyle{eptcs}
\bibliography{epistemicBib}
%
%
\end{document}